\documentclass[twocolumn]{aastex631}

\hypersetup{
    colorlinks=true,
    linkcolor=blue,
    filecolor=magenta,      
    urlcolor=cyan,
   citecolor=blue,
}
\shorttitle{Properties of dark fibrils in the chromosphere}
\shortauthors{Kriginsky et al.}
\graphicspath{{./}{figures/}}
\begin{document}

\title{On the magnetic and thermodynamic properties of dark fibrils in the chromosphere}

\author{Matheus Kriginsky}
\affiliation{Institute for Solar Physics, Dept. of Astronomy, Stockholm University, AlbaNova University Centre, 106 91, Stockholm, Sweden}

\affiliation{Institute of Applied Computing \& Community Code (IAC3) E-07122 Palma de Mallorca, Spain}

\author{Ramon Oliver}

\affiliation{Institute of Applied Computing \& Community Code (IAC3) E-07122 Palma de Mallorca, Spain}
\affiliation{Departament de F\'\i sica, Universitat de les Illes Balears E-07122 Palma de Mallorca, Spain}


\begin{abstract}
Fibrillar structures are ubiquitous in the solar chromosphere and their potential for mediating the mass and energy transport in the solar atmosphere is undeniable. An accurate determination of their properties requires the use of advanced high-resolution observations which are now becoming broadly available from different observatories. We exploit the capabilities of multi-atom, multi-line spectropolarimetric inversions using the Stockholm Inversion Code (STiC). Non-local thermodynamic equilibrium inversions of a fibril-rich area are performed using spectropolarimetric observations in the \ion{Ca}{2} 854.2 nm line obtained with the CRISP imaging spectropolarimeter and spectroscopic observations in the \ion{Ca}{2} H line obtained with the CHROMospheric Imaging Spectrometer (CHROMIS) at the Swedish 1-meter Solar Telescope (SST). Additionally, coobservations in the \ion{Mg}{2} h \& k lines obtained with the Interface Region Imaging Spectrograph (IRIS) are used in the inversions to better constrain the thermodynamic properties of the fibrils. The incorporation of multiple atomic species and spectral lines proves to better constrain the properties of the plasma constituting the fibrils. In particular, the tracing of a large number of fibrils allowed for the study of the variation of the temperature and magnetic field along their projected length over the field of view. The results provide a view of fibrils possessing hot footpoints of about 5 900~K. The temperature drop away from the footpoints is on average 250~K, with a larger drop of around 500~K for the longer fibrils. The magnetic field is also reported to be larger at the footpoints, being almost twice as large as the minimum value reported at the middle point of the fibrils. 

\end{abstract}

%%\keywords{}


\section{Introduction} \label{sec:intro}

One of the most remarkable and common features of the solar chromosphere is the abundance of material resembling dark, slender fibres. These structures are rooted in the magnetic concentrations present in the photosphere, connect regions of opposite polarity and appear to trace the magnetic field as it expands with height. Some appear as more vertical structures which are more easily spotted in observations of the solar limb, and in such a context they are usually called spicules. When observing the solar disc, one of the most easily detectable type of fibrils are those that seem to be more horizontal, and several studies have focused on deciphering their magnetic properties, with a particular interest on whether they are in fact tracing the magnetic field lines \citep[][]{2011A&A...527L...8D,2013ApJ...768..111S,2015SoPh..290.1607S,2017A&A...599A.133A,2017ApJS..229...11J,2022A&A...662A..88V}. This question has also been addressed in parallel with the analysis of numerical simulations \citep[][]{2015ApJ...802..136L,2016ApJ...831L...1M}. The fibrils seem to be mostly following the magnetic field lines, although there are cases where there is a noticeable discrepancy.

The chromospheric nature of fibrils hinders the possibility for the inference of their physical state, as the physical conditions in the chromosphere are complex to model and require the treatment of spectral line formation in the context of non-local thermodynamical equilibrium (non-LTE). Additionally, there is a relative scarcity of lines currently used for diagnostics in the chromosphere when compared to the number of lines that are used for studies of the photosphere or the corona. The majority of works related to chromospheric observations are centred in spectral lines from the hydrogen, helium, singly-ionised calcium and singly-ionised magnesium atoms. Although small in number, these lines have proven to be extremely useful for the determination of the physical properties of the chromosphere. In particular, the numerical studies conducted by \citet{2013ApJ...772...89L,2013ApJ...772...90L} centred on the formation properties of the \ion{Mg}{2} h \& k lines, whose cores form right below the transition region, have shown that they can serve as great temperature diagnostics in the middle to upper chromosphere due to the sensitivity of their emission peak intensities to the temperature of the plasma in the region where they are formed. This conclusion was reached by studying synthetic spectral profiles computed with a 3D radiative transfer code using an atmosphere from a 3D numerical simulation performed with the Bifrost code \citep{2011A&A...531A.154G}. The sensitivity of the \ion{Mg}{2} h \& k lines to temperature has since then been used to constrain the temperature in the chromosphere in a series of works \citep{2016ApJ...830L..30D,2018A&A...620A.124D,2019A&A...627A.101V}. \citet[][]{2018A&A...611A..62B}{}{} performed a similar study, focusing this time in the formation of the \ion{Ca}{2} H \& K lines, which are usually formed lower in the chromosphere than the \ion{Mg}{2} h \& k lines. These authors concluded that the \ion{Ca}{2} H \& K lines can indeed serve as good temperature diagnostics in the middle and lower chromosphere.

With the evolution of the instrumentation, it has become regularly possible to resolve the fine fibrillar structures of the chromosphere. This has allowed for attempts to determine their thermodynamic and magnetic properties using non-LTE inversion codes. \citet[][]{2020A&A...644A..43P}{}{} used spectropolarimetric observations of a fibrillar region near a plage area obtained in the \ion{Ca}{2} 854.2 nm line with the CRISP imaging spectropolarimeter \citep{Scharmer_2008} at the Swedish 1-meter solar telescope \citep{sst} together with spectroscopic observations in the \ion{Ca}{2} K line obtained with the CHROMospheric Imaging Spectrometer \citep[CHROMIS;][]{2017psio.confE..85S}. These authors performed non-LTE inversions of the \ion{Ca}{2} 854.2 nm line using the Stockholm inversion code \citep[STiC;][]{2016ApJ...830L..30D,2019A&A...623A..74D}. The main goal of their work was to study the magnetic field configuration of plage areas, but they also reported inverted temperatures in fibrillar areas of around 4800 K. \citet[][]{2020A&A...637A...1K} studied the so-called \ion{Ca}{2} K bright fibrils using spectroscopic observations in the \ion{Ca}{2} K line, along with spectropolarimetric observations in the \ion{Ca}{2} 854.2~nm and \ion{Fe}{1} 630.2~nm lines. These authors performed non-LTE inversions of the bright fibrillar material, with temperatures of the order of 5000 to 6000~K being reported. These authors did not specifically target the dark fibrillar areas present in their data, as the work was mostly centred in the study of the \ion{Ca}{2} K bright fibrils. \citet{2023A&A...672A..89K} used spectroscopic data obtained with SST/CRISP aligned with observations in the \ion{Mg}{2} h \& k lines obtained with the Interface Region Imaging Spectrograph \citep[IRIS,][]{2014SoPh..289.2733D} whose field of view contained a large number of dark fibrillar material. The aligned observations were inverted with STiC in order to trace the temperature variation along the fibrils. These authors found temperatures at the footpoints of the traced fibrils to be in the range 5000$-$6000~K and on average 300 K larger than the temperature at their midpoints. This decrease in temperature was found for all the traced fibrils. However, the alignment procedure between the CRISP and IRIS observations made it difficult to trace a statistically significant number of fibrils.

Another physical quantity of great relevance in the chromosphere is the magnetic field. While there are very few chromospheric lines whose formation properties are sensitive enough to the magnetic field, the \ion{Ca}{2} 854.2 nm line formed in the infrared has proven to be a line of invaluable usefulness for magnetic field estimations. Its polarisation is sensitive to the magnetic field in the chromosphere and upper photosphere. Forming in the chromospheric environment, it has been extensively used as a magnetic field probe through the computationally cheap weak-field approximation \citep[WFA;][]{1973SoPh...31..299L}. \citet{2020A&A...642A.210M} devised a spatially regularised version of the WFA rooted in the imposition of Tikhonov regularisation in order to infer the magnetic field properties as a function of height in a plage area near the solar disc centre. Their data, obtained with SST/CRISP,  contained several fibrillar structures in the field  of view around the plage area, and the total magnetic field inferred with their method (combining the linear and circular polarisation signals) was found to be up to $10^3$ G in the plage area, and around $10^2$ G in the fibrillar region. Several works have used the formulation of \citet{2020A&A...642A.210M} of the WFA for the determination of magnetic fields, both in active regions \citep{2022A&A...662A..88V} and plage areas \citep{2020A&A...642A.210M,2023ApJ...954L..35D}.

Multi-line inversions help to constrain the properties of the solar atmosphere by introducing additional observational curbs on the inverted physical properties \citep{2018A&A...620A.124D}. \citet{2023A&A...672A..89K} found that the addition of the diagnostic capabilities of the \ion{Mg}{2} h \& k lines to those of the \ion{Ca}{2} 854.2 nm line was crucial in order to infer the thermodynamic properties of the chromospheric fibrils. The temperature sensitivity of the \ion{Ca}{2} K line also aided with the diagnostics performed by \citet[][]{2020A&A...637A...1K}.

The goal of the present work is to exploit the diagnostic capabilities of the \ion{Mg}{2} h \& k lines combined with those of the \ion{Ca}{2} H and \ion{Ca}{2} 854.2 nm lines to infer the thermodynamic and magnetic properties of chromospheric fibrils. The organisation of the work is as follows. Section~\ref{sec:obs} details the different observations used and their calibration, with the inversion scheme and data analysis of spectral features detailed in Sect.~\ref{sec:Data}. The results are discussed in Sect.~\ref{sec:results}, and conclusions are provided in Sect.~\ref{sec:conc}
\section{Observations} \label{sec:obs}

\begin{figure*}
 \centering
 \includegraphics[width=17cm]{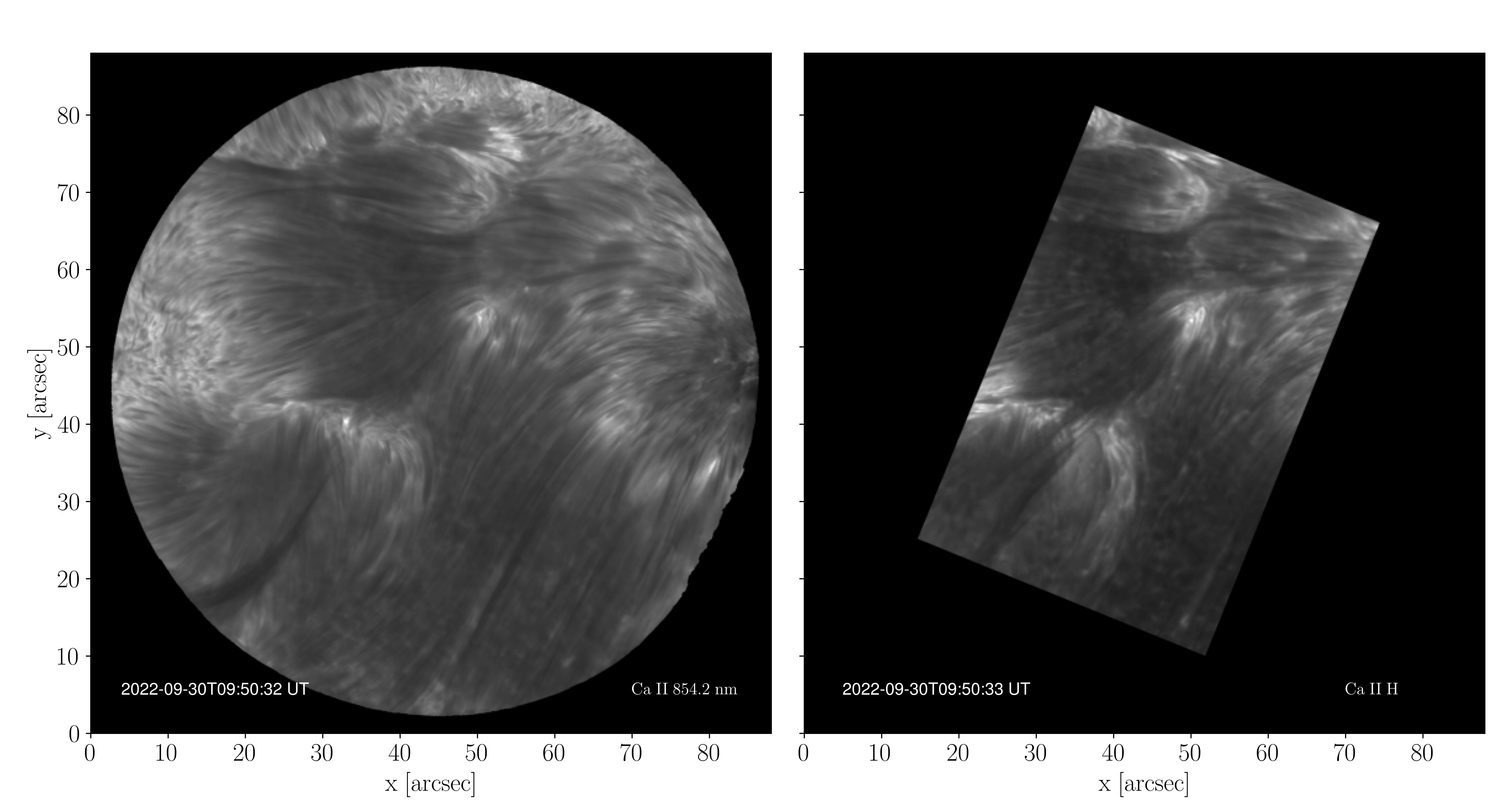}
 \includegraphics[width=17cm]{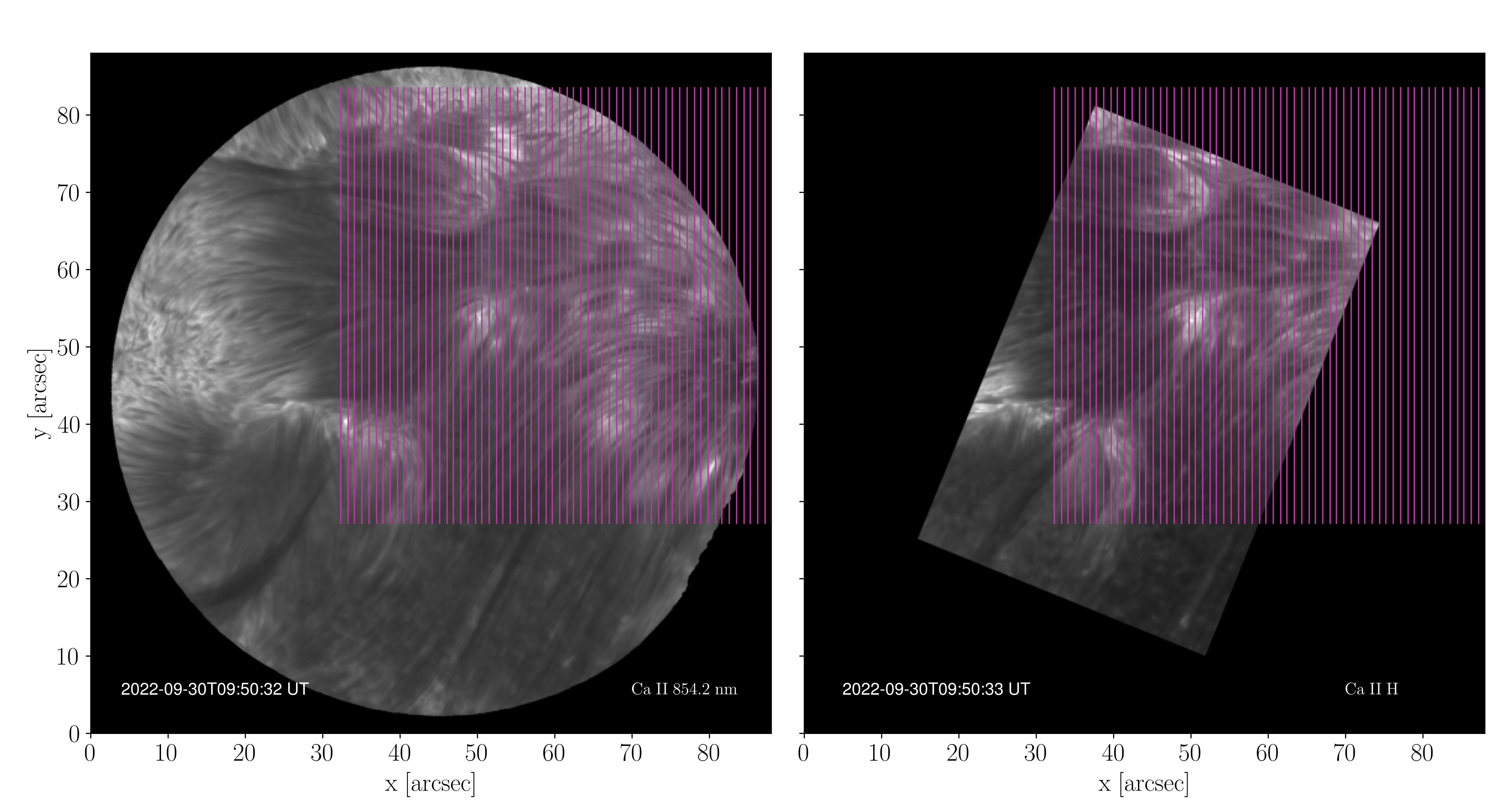}
 \caption{Field of view of the observations. The intensity in the core of the \ion{Ca}{2} 854.2 nm line is shown in the left panels, and the intensity in the core of the \ion{Ca}{2} H line is shown in the right panels. The bottom panels are identical to the ones at the top, with the scanning positions of the slit of IRIS overplotted. The time of the displayed observations is shown in each panel.}
 \label{Figure:1}%
\end{figure*}

The target of the observations used in this work was the region East of the leading sunspot of active region NOAA 13110 as seen on 30 September 2022 (see Fig.~\ref{Figure:1}). Spectroscopic data were obtained with the SST/CHROMIS instrument in the H$\beta$ and \ion{Ca}{2} H lines, with a total cadence of 15 s. The H$\beta$ observations were not used for this work. The \ion{Ca}{2} H line was sampled in 19 positions\footnote{In \AA\hspace{2pt}from the line core: [0, $\pm 0.1$, $\pm 0.2$, $\pm 0.3$, $\pm 0.4$, $\pm 0.5$, $\pm 0.6$, $\pm 0.8$, $\pm 1.0$, $\pm 1.2$, $+ 32$] } around the line core, plus an additional continuum point. At the same time, spectroscopic data in the H$\alpha$ line and spectropolarimetric data in the \ion{Ca}{2} 854.2 nm line were obtained with the SST/CRISP instrument, with a cadence of about 30 s. The H$\alpha$ line data were not used. The \ion{Ca}{2} 854.2 nm line was sampled in 22 positions around the line core position\footnote{In \AA\hspace{2pt}from the line core: [0, $\pm 0.07$, $\pm 0.14$, $\pm 0.21$, $\pm 0.28$, $\pm 0.35$, $\pm 0.45$, $\pm 0.60$, $\pm 0.75$, $\pm 0.95$, $\pm 1.75$, $+ 2$]}. The observations with SST/CRISP and SST/CHROMIS started at 09:45 UT, finishing at 10:15 UT. The total observed frames with CRISP were 57, while for CHROMIS there were 117 completed scans. The data reconstruction was performed using the Multi-Object Multi-Frame Blind Deconvolution \citep[MOMFBD;][]{VanNoort2005} method. Additionally, a new version of the SST data reduction pipeline specifically updated to handle the new cameras employed by CRISP was used before and after the application of the MOMFBD method \citep{2021A&A...653A..68L}. The adaptive optics system used in the SST is described in detail in \citet{2023arXiv231113690S}. An absolute wavelength calibration of the \ion{Ca}{2} H line and the \ion{Ca}{2} 854.2 nm line was done using the atlas profile of \citet[][]{1984SoPh...90..205N}.

At the same time, IRIS was performing medium sparse raster scans over the same region (see Fig.~\ref{Figure:1})  with OBSID 3620258058. The step cadence was around 9~s, with a total of 64 steps per raster scan, each raster scan taking 595~s to complete. A total of three IRIS scans were completed during the time that the observations at the SST were being performed. The data from the IRIS observations were calibrated following the reduction methods detailed in \citet[][]{2018SoPh..293..149W}. Additionally, radiometric calibrations were carried out using version four of the calibration files used by the
\textit{iris\_get\_response} routine in SolarSoft \citep[SSW;][]{1998SoPh..182..497F}.
 
 \begin{figure*}
 \centering
 \includegraphics[width=17cm]{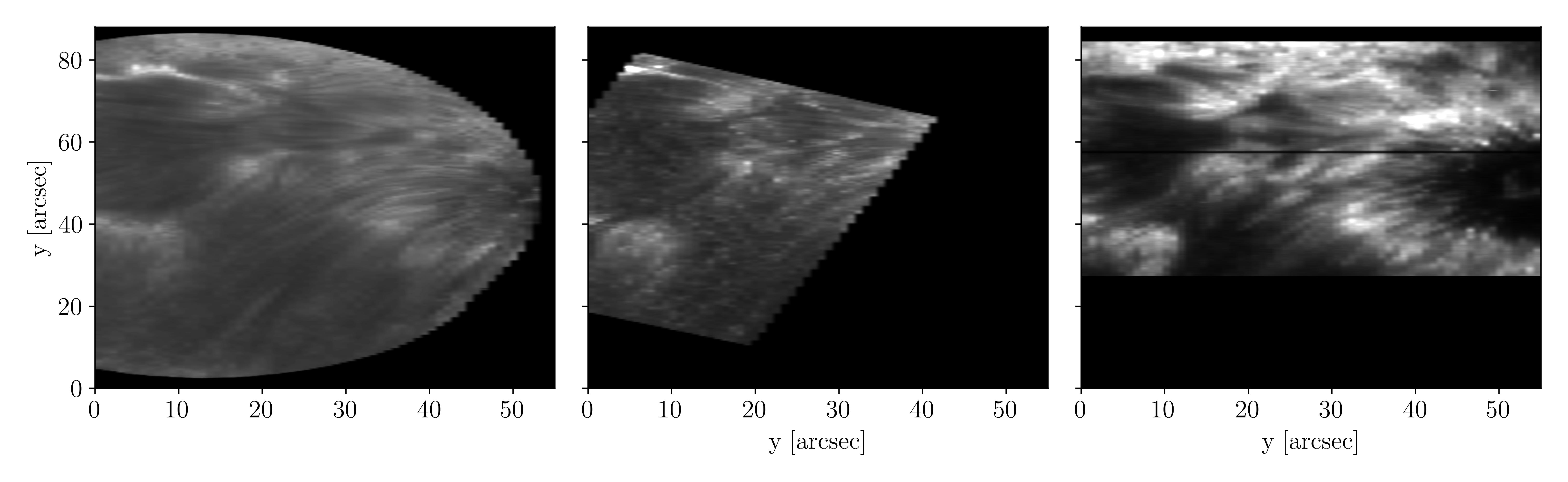}
 \includegraphics[width=17cm]{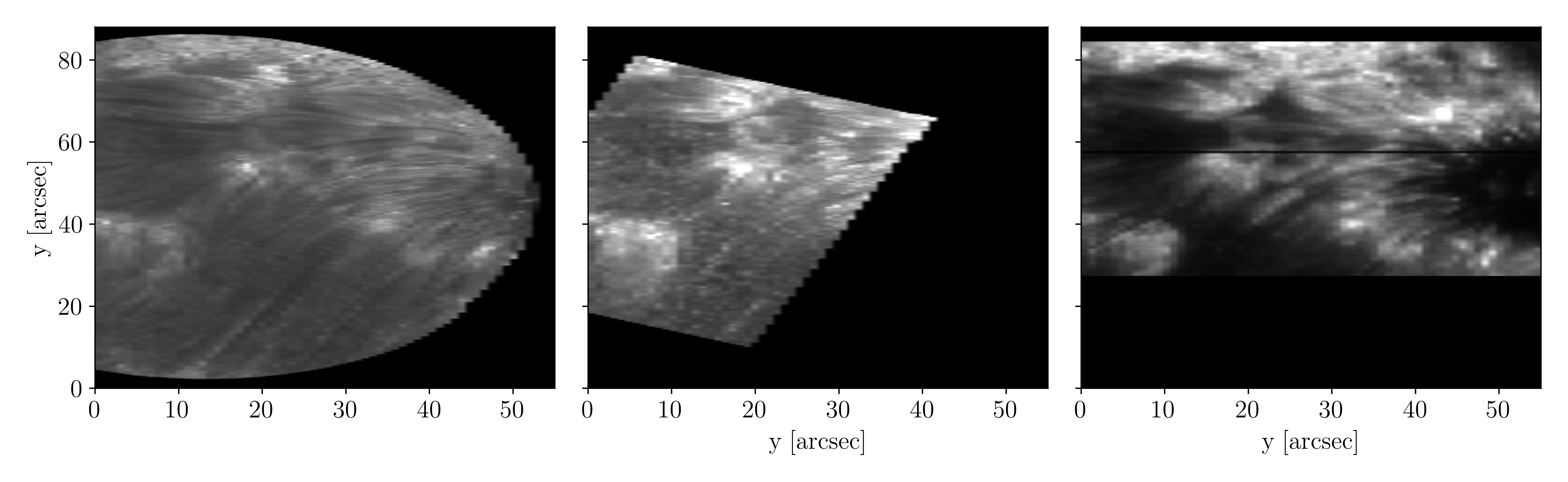}
 \includegraphics[width=17cm]{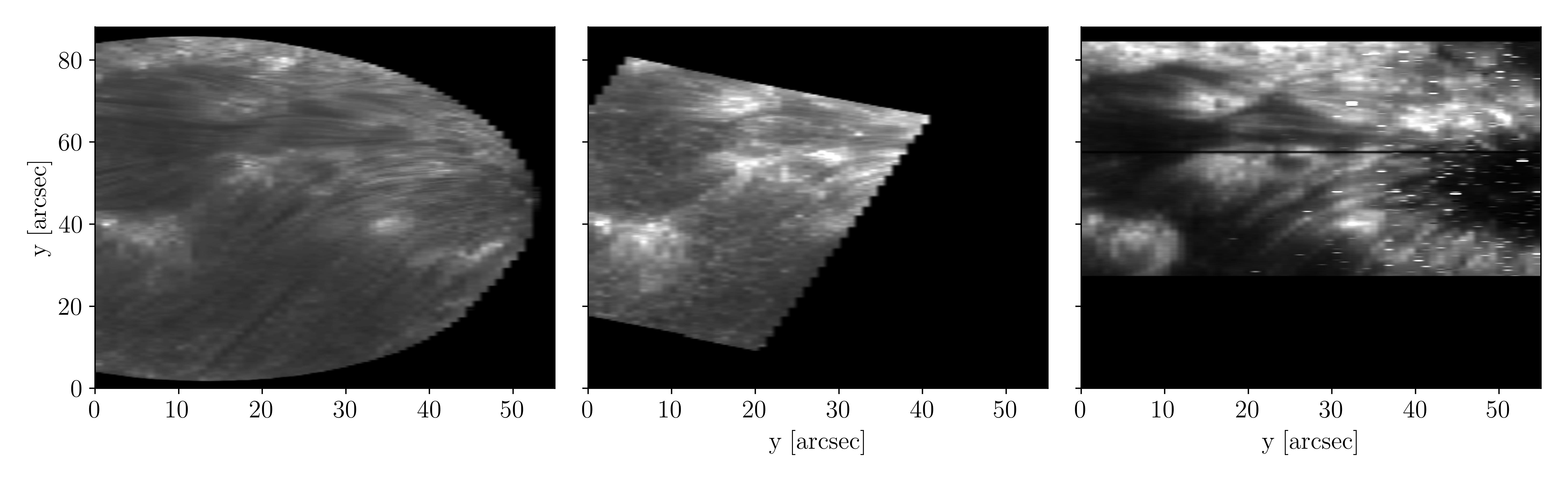}

 \caption{Coaligned datasets. In each row, the three different coalined datasets corresponding to each raster scan of the IRIS observations are displayed. The first, second and third columns display the intensity in the cores of the \ion{Ca}{2} 854.2~nm line, the \ion{Ca}{2}~H line and the \ion{Mg}{2}~k line, respectively. The bright pixels on the bottom right panel correspond to bad pixels probably caused by particles hitting the detector at the time of the observations.}
 \label{Figure:2}%
\end{figure*}

\subsection{Data alignment}
The new CRISP cameras perform scans with a pixel sampling size of 0.044\arcsec\ in the \ion{Ca}{2} 854.2 nm line. This new pixel sampling size is closer to that of CHROMIS in the \ion{Ca}{2} H line, namely 0.038\arcsec. Therefore, the choice was made to upsample the original CRISP observations to match the pixel size of CHROMIS. For each of the 57 scans performed by CRISP, the corresponding CHROMIS scan closest in time to it was selected, and the images from both instruments were coaligned, totalling 57 scans containing cospatial, nearly cotemporal (the maximum time difference between both observations was 7.5 s) observations in both \ion{Ca}{2} lines. The remaining 60 CHROMIS scans were not used.

The alignment procedure of the SST observations with the IRIS raster scans was performed identically to the methodolody described by \citet{2023A&A...672A..89K}, ending up with a coaligned data series where the \ion{Ca}{2} data was averaged spatially in order to match the spatial resolution of the IRIS observations. The only difference in the final result is that the IRIS observations used in this work have half the pixel sampling size (0.16\arcsec) of the observations used by these authors (0.33\arcsec). The three resulting coaligned raster scans are displayed in Fig.~\ref{Figure:2}. The maximum time difference between the \ion{Ca}{2} and \ion{Mg}{2} observations after alignment was 11~s. It is important to note that the horizontal direction of the images has only 62 pixels, compared to the 550 pixels in the vertical direction. This is caused by the non-contiguous sampling of the raster scans (see purple lines in the bottom row of Fig~\ref{Figure:1}). There are effectively four pixels "missing" between each pair of consecutive pixels in the horizontal direction.

\section{Methods}\label{sec:Data}

 \subsection{Inversions} 
 \subsubsection{STiC}
 In order to determine the temperature at the formation height of the fibrils, non-LTE inversions of the spectroscopic data were performed using the STiC inversion code, a regularised Levenberg-Marquardt code that uses an optimised version of the RH code \citep[][]{2001ApJ...557..389U} to solve the atom population densities assuming statistical equililibrium and plane-parallel geometry for multi-atom inversions of multiple spectral lines. STiC can account for important effects in the formation of the \ion{Ca}{2} H and \ion{Mg}{2} h \& k lines such as partial redistribution of scattered photons \citep[][]{2012A&A...543A.109L}. Cubic Bezier solvers are used to solve the radiative transfer equation \citep[][]{2013ApJ...764...33D}.
 STiC includes an equation of state extracted from the Spectroscopy Made Easy code \citep[SME;][]{2017A&A...597A..16P}. It is important to note that STiC is a 1.5D code, therefore some 3D effects that are relevant to the formation of the line cores of the \ion{Ca}{2} H and \ion{Mg}{2} h \& k lines cannot be completely accounted for.

 \subsubsection{Model atoms}

The \ion{Mg}{2} atom was modelled using 11 atomic levels including the ground level of \ion{Mg}{3}. Computation time was saved by using only the spectral region near the cores of the \ion{Mg}{2} h \& k lines and the region near the two triplet lines between them for the inversions, employing the same spectral regions as those used by \citet[][see their Fig.~4]{2023A&A...672A..89K}. A model atom for \ion{Ca}{2} with five levels plus the continuum was used. The \ion{Ca}{2} 854.2 nm line was modelled using a Voigt profile, while the \ion{Mg}{2} h \& k and \ion{Ca}{2} H lines were treated accounting for the effects of partial frequency redistribution.

 \subsubsection{Input atmosphere}
 
 While in the overwhelming majority of works that include inversions the choice of the depth variable has been the optical depth $\tau$, STiC also offers the possibility to work with the column mass density $\xi$ as the depth variable. Not only is the choice to use $\xi$ better for the hydrostatic equilibrium consideration from a numerical point of view \citep{2019A&A...623A..74D}, it also has advantages over $\log \tau$ when the inversions are performed in lines that form in the chromosphere, since the $\tau$ scale compresses the chromosphere, while it is much more extensive in a $\log \xi$ scale \citep{2019A&A...623A..74D,hoffmann2022}.

 A regular depth grid of 35 points covering the range $\log \xi = [-4.13,1]$\footnote{$\xi$ has units of g $\mathrm{cm}^{-2}$.} was used in the inversions. The initial guess atmosphere was an interpolated FAL-C \citep[][]{1985cdm..proc...67A,1993ApJ...406..319F} model. The parameters that were actively inverted were the temperature, $T$, the line-of-sight component of the velocity, $v_{\mathrm{LoS}}$, and the microturbulent velocity, $v_{\mathrm{turb}}$.

\begin{figure*}
 \centering
 \includegraphics[width=17cm]{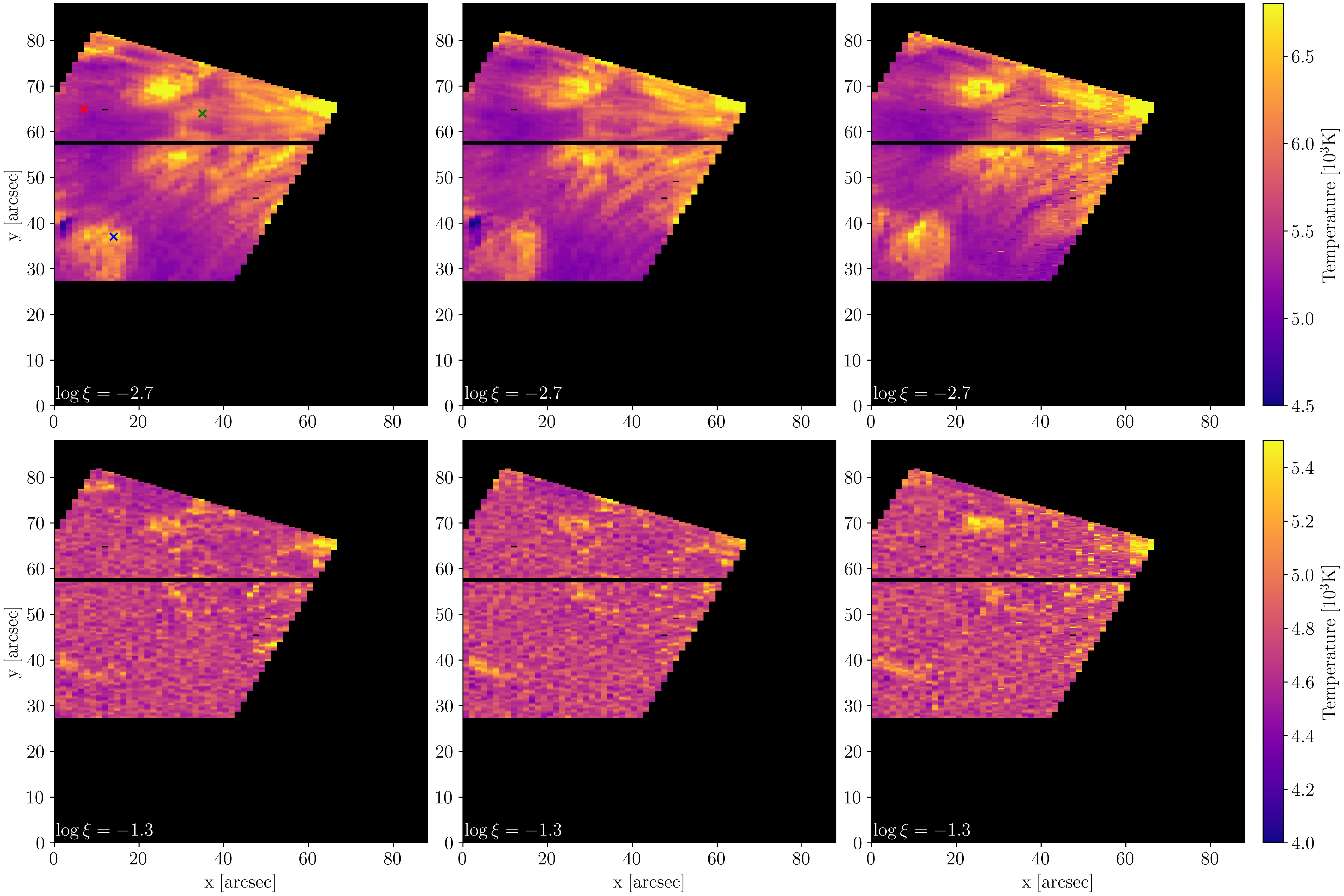}

 \caption{Temperature inversions. The temperature inversions in the chromosphere are shown in the top row, with the temperature in the photosphere shown in the bottom row. Each column represents one of the three coaligned observations between CRISP, CHROMIS and IRIS (shown in Fig.~\ref{Figure:2}). The column mass of each temperature map is written in each image. The pixels marked in the top left panel pinpoint the locations used for the curves in Fig.~\ref{Figure:4}.}
 \label{Figure:3}%
\end{figure*}

\begin{figure}
 \centering
 \includegraphics[width=8cm]{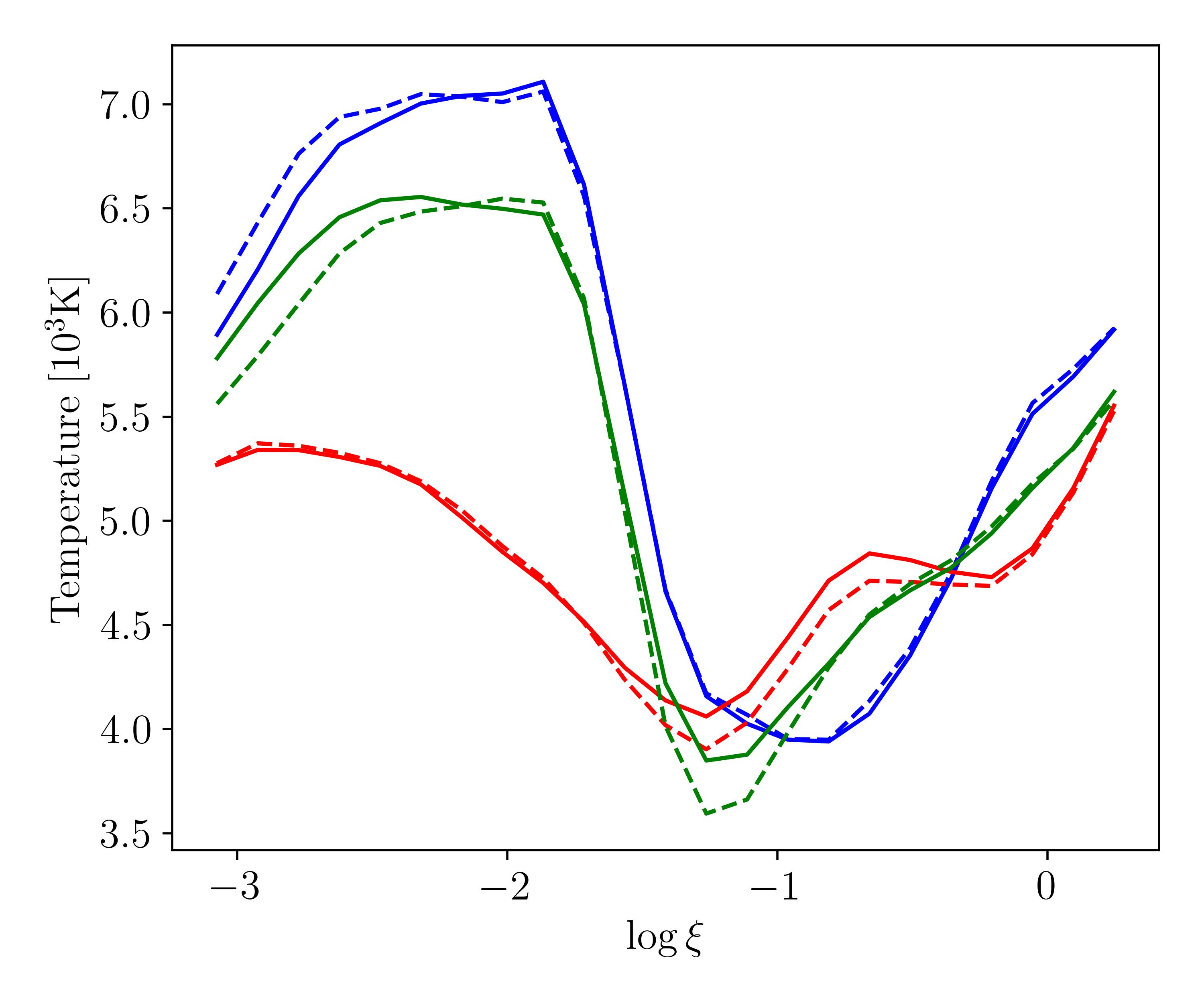}

 \caption{Error in the enhanced inversions. The stratification of the temperature is shown for three pixels marked in the same colours on Fig.~\ref{Figure:3}. The continuous lines represent the enhanced inversions performed with the NN. The dashed lines show the stratification of temperature obtained from inversions with STiC including all spectral lines.}
 \label{Figure:4}%
\end{figure}

 \subsubsection{Inversion strategy}
 The large number of pixels (over 150 million) in the observations made it computationally unattainable to invert every single one of them with STiC. We therefore devised a strategy to incorporate neural networks (NNs) to speed up the inversion process. We selected five equidistant non-consecutive scans from the original 57 of the coaligned datasets containing the \ion{Ca}{2} H and \ion{Ca}{2} 854.2 nm lines. For each scan, every fifth pixel was inverted with STiC using a FAL-C atmosphere as input. The results of the inversions were used to train three 1D NNs identical in architecture to each other that would predict the stratification of each of the actively inverted physical parameters ($T$, $v_{\mathrm{LoS}}$ and $v_{\mathrm{turb}}$) independently from each other. The NNs had three hidden dense layers of 150 neurons each, in addition to the input layer (the \ion{Ca}{2} H and \ion{Ca}{2} 854.2 nm line profiles) and output layer (the stratification of $T$, $v_{\mathrm{LoS}}$ and $v_{\mathrm{turb}}$). ReLU activation functions were used for all the dense layers, and the Adam optimiser was employed. The loss function used was the mean squared error. The networks were trained until the loss was below the $10^{-5}$ threshold. The calculated loss for the test data was below $10^{-3}$. Separating the inverted physical parameters into three different NNs was done because the convergence was much faster than using all three parameters as the input of a single NN. It is a similar approach to the one successfully followed by \citet{asensio2019}. 

 With a trained network able to predict the stratification of physical parameters given the \ion{Ca}{2} H and \ion{Ca}{2} 854.2 nm line profiles as input (hereafter referred to as $NN_{\mathrm{Ca}}$), the next step was to perform the inversion with STiC of the coaligned \ion{Ca}{2} and \ion{Mg}{2} data shown in Fig.~\ref{Figure:2}. The input atmosphere guess was the predicted atmosphere with $NN_{\mathrm{Ca}}$, and all the pixels of the coaligned data were inverted with STiC.

 \subsection{Enhanced inversions}

 The inversions of the coaligned \ion{Ca}{2} and \ion{Mg}{2} data performed with STiC, while successful in reproducing the observed spectral lines, could only be performed in the overlapping regions of the field of view of the three instruments, both spatially and temporally. The field of view of the original CRISP observations in the \ion{Ca}{2} 854.2 nm line shown in Fig.~\ref{Figure:1} contains a much larger variety of fibrils.

 A way of using the full field of view of the CRISP observations would be to perform inversions of the \ion{Ca}{2} 854.2 nm line alone since it is sensitive to temperature both in the photosphere and the chromosphere. However, this would undermine the additional constraints of multi-line, multi-atom inversions. \citet{2023A&A...672A..89K} showed how the use of the \ion{Ca}{2} 854.2 nm line alone can fail to reproduce the fine thermodynamic properties of the fibrillar material. 

 To circumvent the issues presented in the previous paragraphs, a scheme to use the \ion{Ca}{2} 854.2 nm line alone while maintaining the improved constraints of the multi-line inversions was devised. The first step was to use the inversion results of the coaligned \ion{Ca}{2} and \ion{Mg}{2} observations to determine the response functions to temperature of the \ion{Ca}{2} 854.2 nm line. We found that there is, on average, a significant response to temperature in the depth range $\log \xi$ = [$-3, 0$]. We then trained a neural network to connect the observed \ion{Ca}{2} 854.2 nm spectral profile of each pixel in the coaligned data with the temperature stratification over this depth range. The role of this NN is to provide enhanced inversions of the \ion{Ca}{2} 854.2 nm line at depths where this spectral line is sensitive to temperature but perhaps not sensitive enough to lead a regular inversion code like STiC to the refined temperature stratification that can only be achieved by introducing the constraints of additional spectral profiles. This 1D NN had 4 hidden dense layers of 800 neurons each, in addition to the input and output layers. ReLU activation functions were used, and the Adam optimised was chosen. The NN was trained in 90 \% of the coaligned data, yielding a validation loss lower than $5 \times 10^{-4}$.
\begin{figure*}
 \centering
 \includegraphics[width=17cm]{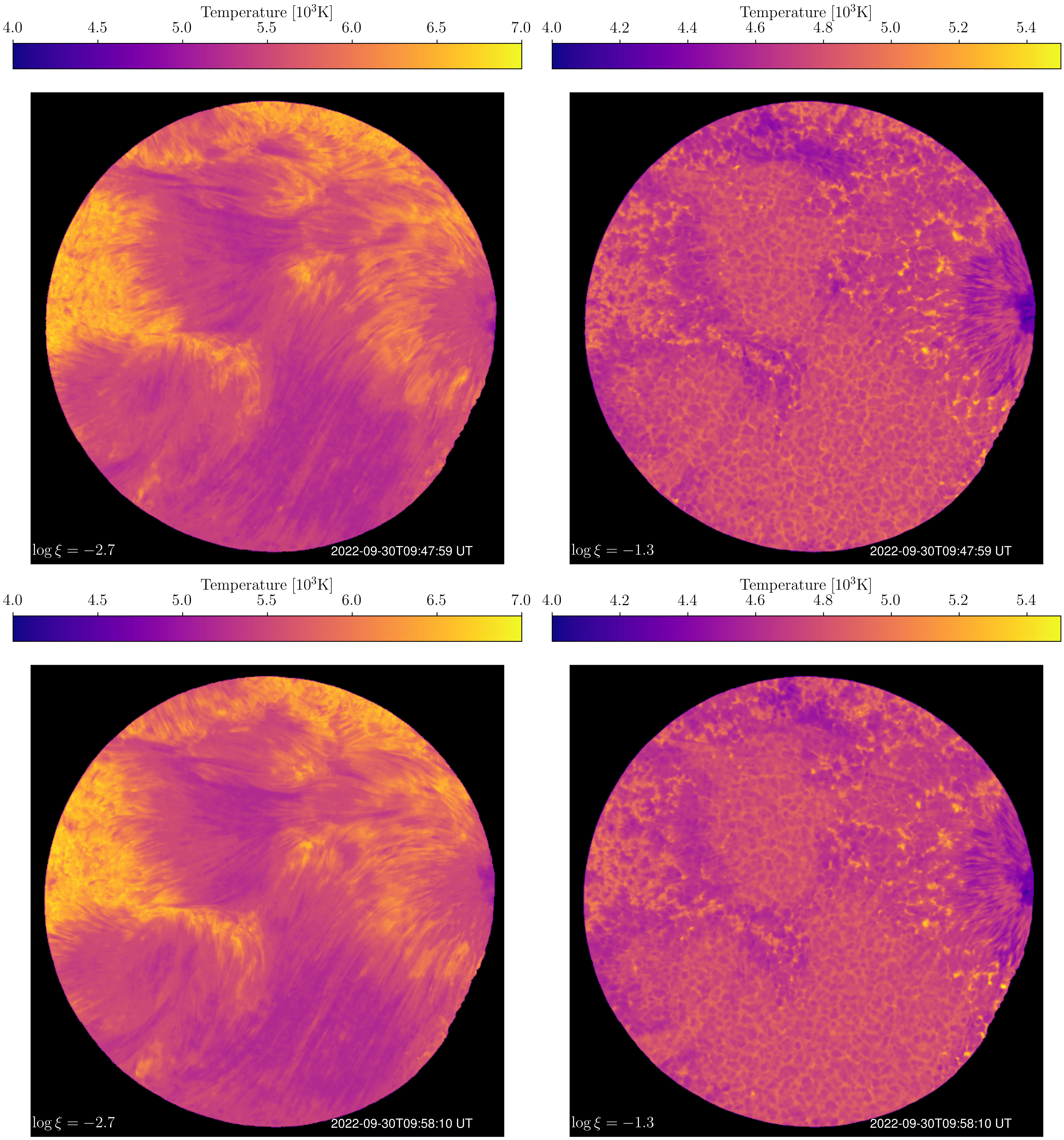}

 \caption{Temperature inversions. Temperature inversions in the chromosphere for two different scans are shown in the left column, with the corresponding temperature maps in the photosphere shown in the right. The column mass of each temperature map is written in each image.}
 \label{Figure:5}%
\end{figure*}
 While in practice it is feasible to perform a similar correction to other actively inverted parameters such as $v_{\mathrm{LoS}}$ and $v_{\mathrm{turb}}$, we decided to refrain from doing so as these parameters are usually inverted with the use of fewer nodes in regular inversions when compared to the nodes used for the temperature stratification. Additionally, it might prove challenging to assess the usually troublesome ambiguity that comes with inferring the microturbulent velocity in single-atom inversions.

 \subsection{Magnetic field diagnostics}\label{sect:43}

While in principle STiC allows for the inversion of the magnetic field vector, we chose to reduce the complexity of the inversions by not including the spectropolarimetric data from the  \ion{Ca}{2} 854.2 nm line observations. Instead, the formulation of the WFA with spatial regularisation introduced by \citet{2020A&A...642A.210M} was used to infer the magnetic fields in the chromosphere. 

The \ion{Ca}{2} 854.2 nm line is sensitive to magnetic fields in the chromosphere (around the line core) and in the photosphere (at the line wings). Therefore, we only used the region within the range [$-0.025$, $0.025$] nm from the line core for the determination of the line-of-sight component of the magnetic field ($B_{\mathrm{LoS}}$). The linear polarisation signals from the \ion{Ca}{2} 854.2 nm line also have sensitivity to magnetic fields both in the chromosphere and in the photosphere, and their signals are used to determine the plane-of-the-sky component of the magnetic field ($B_{\mathrm{PoS}}$). However, the formulation employed by \citet{2020A&A...642A.210M} is only valid away from the line centre. Therefore, we chose to use the same spectral region employed by \citet{2017A&A...599A.133A}, namely the region  [$-0.035, -0.014$]~$\cup$~[$0.014$, $0.035$] nm from the line centre. This spectral range still has enough signal in Stokes $Q$ and $U$ while being far from the line core. Additionally, it still has significant sensitivity to the magnetic field in the chromosphere \citep{2016MNRAS.459.3363Q}.

\section{Results} \label{sec:results}

\subsection{Inverted temperatures}
The inverted temperature maps for each of the three coaligned datasets of Fig.~\ref{Figure:2} are shown in Fig.~\ref{Figure:3}. The chosen $\log{\xi}$ values are the ones where the temperature morphology most closely resembles that of the fibrils ($\log{\xi} =$~$-2.7$) and the photosphere ($\log{\xi} =$~$-1.3$) seen in the observations. The scattered bright and dark points present mostly in the third column of Fig.~\ref{Figure:3} are the pixels where the data acquisition of the IRIS raster failed, likely due to particles hitting the detector during the observations. The black horizontal lines represent the same horizontal lines from the \ion{Mg}{2} h \& k observations, also seen in the third column of Fig.~\ref{Figure:2}. The data from that region were not inverted. The inverted temperature values around the fibrilar areas closely match those reported by \citet{2023A&A...672A..89K}, with an average value around 5 500~K. 

\subsection{Enhanced Ca II inversions}

 To assess the quality of the enhanced inversions obtained by using the NN, the temperature stratification for three different pixels is shown in Fig.~\ref{Figure:4}. The pixels were deliberately chosen in different areas of the field of view: a pixel in the fibril area (in red), a pixel in a plage area (in blue) and a pixel in a region voided of fibrils (in green). For all the pixels not used in the training of the NN, the average absolute error was slightly smaller than 100 K and the maximum error was of 250 K. These discrepancies are small enough to encourage us to use the NN in the remainder of the CRISP field of view.

 Two examples of the temperature maps obtained at $\log{\xi} =$~$-2.7$ and $\log{\xi} = -1.3$ are shown in Fig.~\ref{Figure:5}. The temperature structure morphologically reproduces the fibrillar structures in the chromosphere, while in the photosphere there is a strong correlation with the granulation patterns and the plage and sunspot areas. While the sunspot results are included, the validity of the temperature stratification in these structures is not in principle reliable since no such pixels were used in the training of the NN. As expected, the warmer regions are in the plage areas, where temperatures of the order of 7~000 K can be found. It is also clear that the footpoints where the fibrillar regions are rooted are warmer than the rest of the fibrils themselves.

\begin{figure}[ht]
 \centering
 \includegraphics[width=7cm]{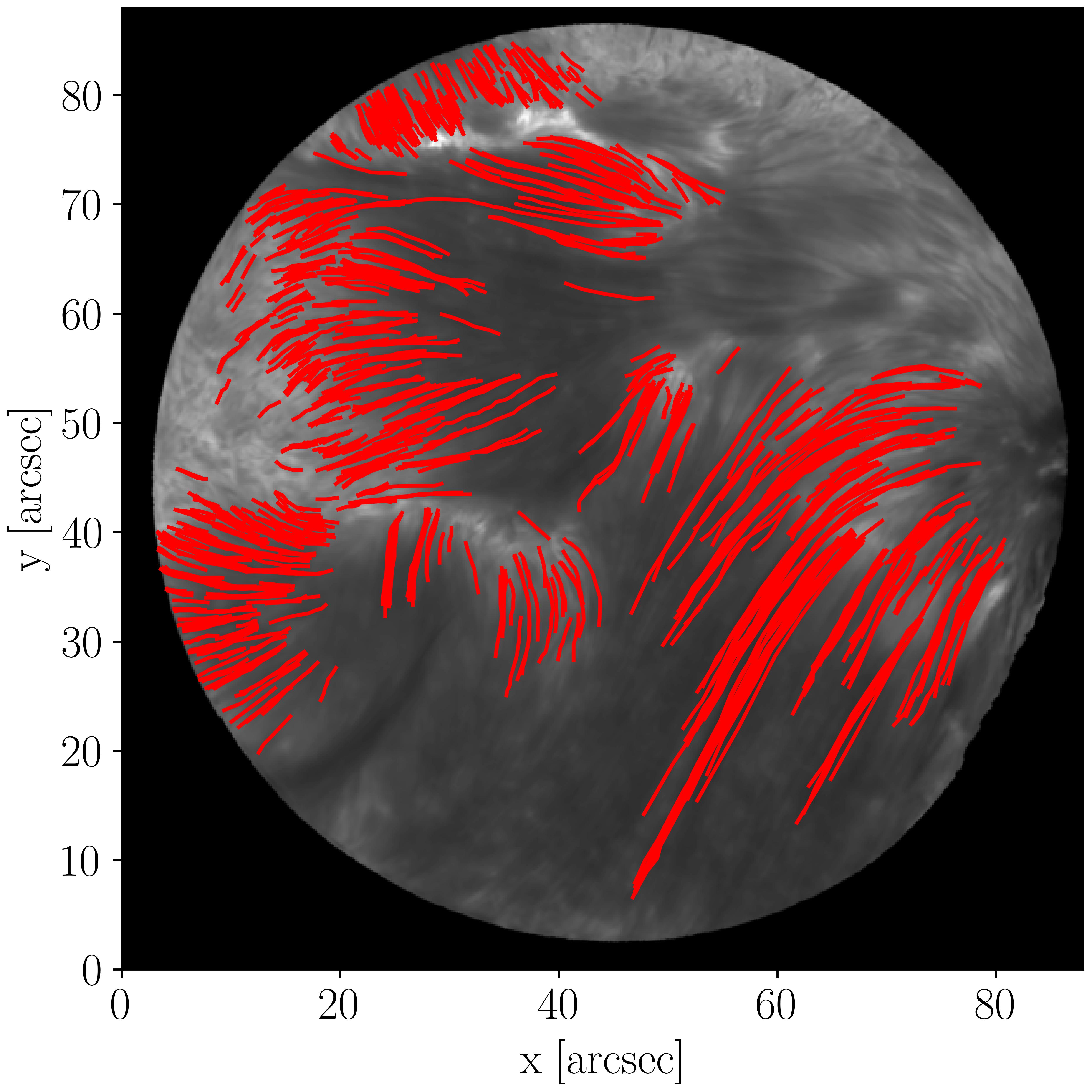}
 \caption{Fibril tracing. The outline of all the traced fibrils is shown as red curves over a \ion{Ca}{2} 854.2 nm line core image. }
 \label{Figure:6}%
\end{figure}
\begin{figure}[ht]
 \centering
 \includegraphics[width=8cm]{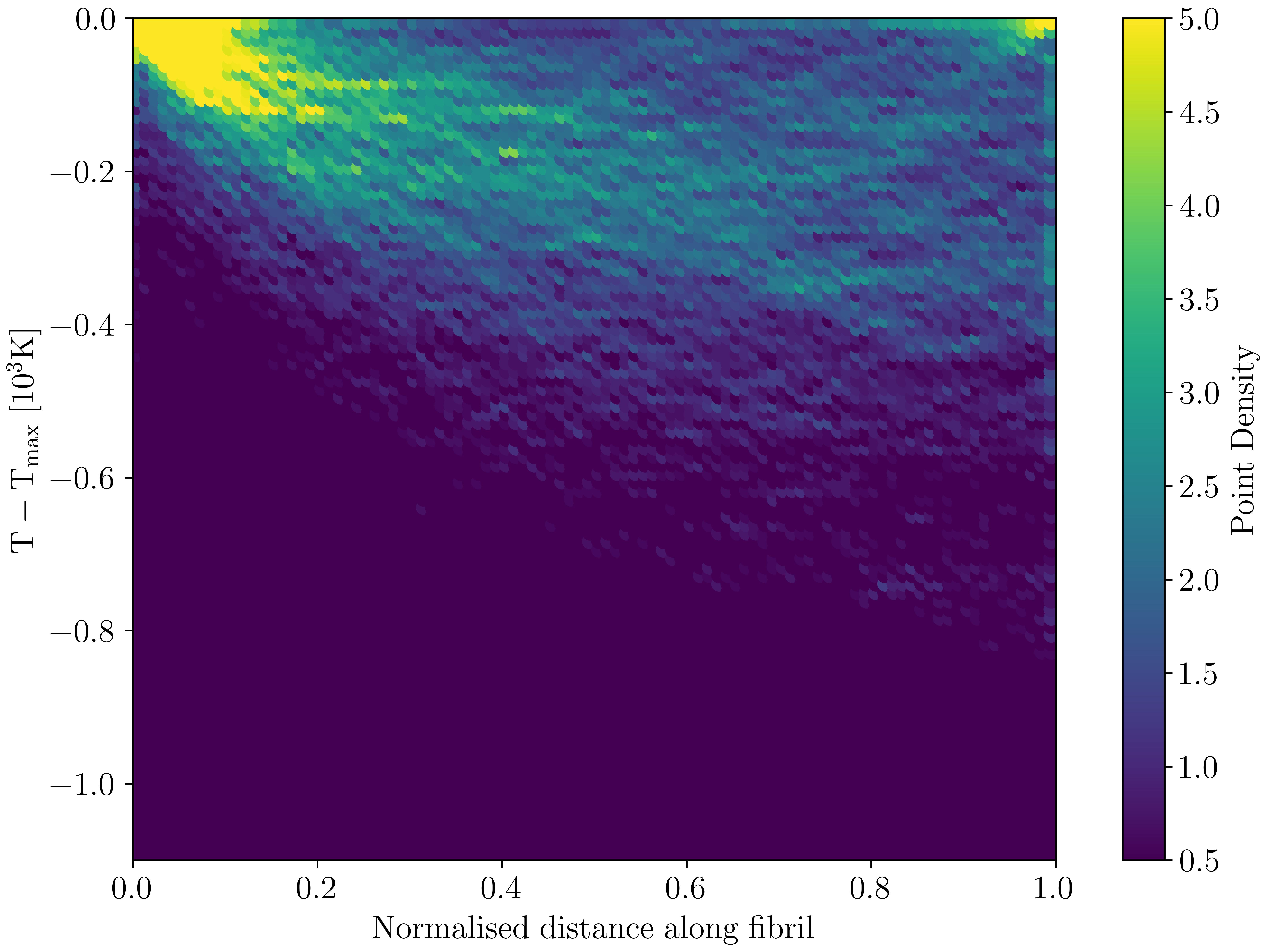}
 \caption{Temperature drop. A heat map of the variation of the temperature over the normalised length of the fibrils, with the leftmost point corresponding to the starting footpoint for the tracing of each individual fibril.}
 \label{Figure:7}%
\end{figure}
\begin{figure*}[ht]
 \centering
 \includegraphics[width=16cm]{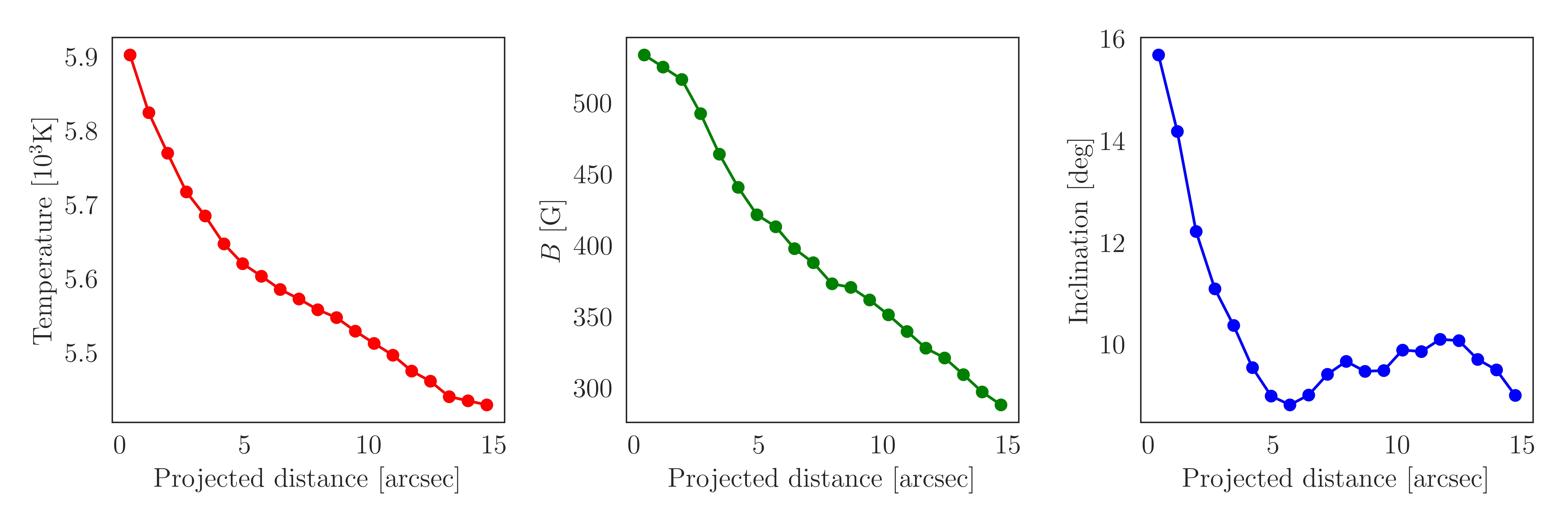}
 \caption{Average variation of the temperature (left),  the magnetic field$|B|$ (middle) and the magnetic field inclination (right) along the fibrils as a function of the projected distance. }
 \label{Figure:8}%
\end{figure*}
 \subsection{Fibril tracing}
Since we were able to determine the temperature structure in the full field of view of the CRISP observations, it was possible to trace a very large number of dark fibrils. We selected 10 temporal scans, tracing a total of 1255 fibrils manually. The totality of the traced fibrils is shown in Fig.~\ref{Figure:6}. The large number of traced fibrils allowed us to generate a heatmap of the variation of the temperature along the projection on the plane of the sky of the fibrils, which is shown in Fig.~\ref{Figure:7}. We normalised the distance along the different fibrils along their visible length in order to plot them together. The temperature is plotted as a temperature difference with respect to the maximum temperature found at the footpoints.

Similarly to \citet{2023A&A...672A..89K}, we found two main types of fibrils. Those that we were able to trace between both footpoints \citep[referred to as group 1 by][]{2023A&A...672A..89K}, which can be seen in the region centred at $x,y = (10,35)$ in Fig.~\ref{Figure:6}, and those whose tracing started at one footpoint but ended in a location where the fibrils could not be individually traced anymore (group 2). It is important to note that these variations are seen on the projected fibrils, as the 3D geometry of the fibrillar areas could not be established from the observations.

The group 1 and group 2 fibrils are clearly distinguishable in Fig.~\ref{Figure:7}. The fibrils belonging to group 1 all had a common temperature variation along their projected length, with a peak temperature at the footpoints, decreasing to a common minimum that on average was 250~K colder than the footpoints and located at the midpoint. The fibrils from group 2 also had a hot footpoint, and their temperature decreased along their length until the point where they are no longer detectable. Group~2 fibrils are also generally longer, and the temperature reached at the tail where they become no longer traceable can be up to 600~K colder than their footpoints. 

We defined 20 bins of 0.75\arcsec\ in size, and computed the average inverted temperature, the magnetic field $|B|$ and its inclination for all the fibrils combined, shown in Fig.~\ref{Figure:8}. The inclination of the magnetic field is defined as

\begin{equation}
    \theta = \arcsin \Bigg( \Bigg| \frac{B_{\mathrm{LoS}}}{B} \Bigg|\Bigg).
\end{equation}

 We chose to use 20 bins because that led to a maximum length of 15\arcsec, which corresponds to the location of the average projected midpoint of the group 1 fibrils. The average temperature at the footpoint of the fibrils is 5~900~K, dropping to 5400~K at a projected distance of 15\arcsec, corresponding to the longest traced fibrils. This behaviour is in good agreement with the results obtained by \citet{2023A&A...672A..89K}. 

 The average value of $|B|$ at the fibril footpoints is around 550 G, with a drop to 280 G at a projected distance of 15\arcsec.

All three variables plotted in Fig.~\ref{Figure:8} show at least two clear stages of their variation. In the first 5\arcsec, the value of the inclination of the magnetic field decreases much more drastically, followed by a more or less constant value around $9^\circ$. A similar behaviour can be seen for the temperature. The reduction in the value of $|B|$ is linear in both regions but it does seem to be a steeper decrease in the first 5\arcsec. Keeping in mind the possible projection effects, this seemingly linear variation of $|B|$ appears to indicate that the dark fibrils studied here host a less complex magnetic structure than that of other jets in the chromosphere, with this linear variation a possible sign of the flux conservation as the cross-sectional area increases with height. One possible way to corroborate this would be to try to measure an expansion of the diameter of the fibrils and compare its variation to the trend shown by the magnetic field. However, with the data at hand, an accurate measurement of the diameter of the fibrils could not be obtained.

Additionally, if we assume that the fibrils studied here are already in an environment where the magnetic pressure is larger than the gas pressure and therefore the structures are aligned with the magnetic field, it appears that the traced fibrils are all very inclined with respect to the vertical direction. This assumption would mean that these dark fibrils are in an environment where the magnetic field has already expanded and is largely horizontal. 

Furthermore, the assumption of the alignment of the plasma in these dark fibrils has implications regarding their footpoint properties. One would imagine the "true" footpoints as rooted in the photospheric magnetic field concentrations with the magnetic fields nearly vertical. The fact that the magnetic field even at the footpoints is highly inclined means that either the magnetic field becomes horizontal very quickly or that the "real" footpoints are not detected here, but only some location in the fibrils close to the footpoints, with the true footpoins being undetectable either due to the resolution, superposition or physical properties of the plasma.

\section{Conclusions} \label{sec:conc}

Constant advances in the instrumentation and telescopes allow us to continuously study smaller structures hosted by the solar atmosphere. However, such advancements come at the price of increasingly larger datasets. While the use of inversions as a diagnostic tool has been performed for decades, the increase in the size of the datasets hinders our ability to perform inversions in all the pixels from a single observation. On this study, we have made use of an increasingly popular technique, namely the use of neural networks. Neural networks trained to perform inversions in a small yet significant portion of a dataset can make the inversion problem several orders of magnitude faster, since an inversion of a pixel with STiC can take around 600~s when inverting all the \ion{Mg}{2} and \ion{Ca}{2} lines, while it takes about $10^{-4}$~s seconds for the NNs to yield a prediction.

While the introduction of neural networks into the world of inversions is a promising prospect for the analysis of large datasets, there are additional challenges that they can help us overcome. It is common to make compromises when combining observations of different instruments and telescopes of different spatial resolution. We have seen how much data can be lost in trying to align and spatially and temporally match observations from a filtergraph and a spectrograph. With an innovative use of neural networks, we have been able to study the observations in the entire field of view of the original CRISP observations without losing the additional diagnostic capabilities of the \ion{Mg}{2} h \& k lines. This was achieved by carrying out enhanced inversions, exploiting inversion results obtained with STiC using multi-line, multi-atom observations that better constrain the thermal stratification of the solar atmosphere. The enhanced inversion technique aided by neural networks was designed to amplify the sensitivity to temperature of the \ion{Ca}{2} 854.2 nm line, as regular inversions would not have yielded the same temperature stratification. This method could be a promising idea, however, it should be noted that it still relies on performing costly inversions of a statistically representative set of pixels so that the neural network can properly reproduce the results of the inversions.

 With the NN-enhanced inversions, we were able to trace 1255 fibrils in the neighbourhood of an active region and a plage area. The inferred temperature values on the fibrillar areas are in great agreement with the work of \citet{2023A&A...672A..89K}, but with a much larger sample of fibrils. We have corroborated the general variation of the temperature along the fibrils, with hot footpoints of around 6000~K, and values at the midpoints on average 250 K colder. Spectropolarimetric observations in the \ion{Ca}{2} 854.2 nm line allowed us to estimate the chromospheric magnetic field using the WFA. The inferred variation of $|B|$ along the projected length of the fibrils is almost linear.

\begin{acknowledgements}
This publication is part of the R+D+i projects PID2020-112791GB-I00 and PID2023-147708NB-I00, financed by MCIN/AEI/10.13039/501100011033. MK acknowledges the support from the Vicepresidència i Conselleria d’Innovació, Recerca i Turisme del Govern de les Illes Balears and the Fons Social Europeu 2014-2020 de les Illes Balears. IRIS is a NASA small explorer mission developed and operated by LMSAL with mission operations executed at NASA Ames Research center and major contributions to downlink communications funded by ESA and the Norwegian Space Centre. The Swedish 1-m Solar Telescope is operated on the island of La Palma by the Institute for Solar Physics of Stockholm University in the Spanish Observatorio del Roque de los Muchachos of the Instituto de Astrofísica de Canarias. The Institute for Solar Physics is supported by a grant for research infrastructures of national importance from the Swedish Research Council (registration number 2017-00625).  
\end{acknowledgements}


%\vspace{5mm}
%\facilities{HST(STIS), Swift(XRT and UVOT), AAVSO, CTIO:1.3m,




\bibliography{sample631}{}
\bibliographystyle{aasjournal}



\end{document}